# Significant enhancement of irreversibility field in clean-limit bulk MgB$_2$

V. Braccini[1,2], L.D. Cooley[1], S. Patnaik[1], P. Manfrinetti[3], A. Palenzona[3], A.S. Siri[2], and D.C. Larbalestier[1]

[1] Applied Superconductivity Center, University of Wisconsin, 1500 Engineering Drive, Madison, Wisconsin 53706
[2] I.N.F.M.-LAMIA, Dipartimento di Fisica, Via Dodecaneso 33, 16146 Genova, Italy
[3] I.N.F.M.-LAMIA, Dipartimento di Chimica e Chimica Industriale, Via Dodecaneso 31, 16146 Genova, Italy

Low resistivity MgB$_2$ bulk samples annealed in Mg vapor show an increase in irreversibility field $H^*(T)$ by a factor of ~2 in both transport and magnetic measurements. The best sample displayed $\mu_0 H^* > 14$ T at 4.2 K and ~6 T at 20 K. These changes were accompanied by an increase of the 40 K resistivity from 1.0 to 18 $\mu\Omega$cm and a lowering of the resistivity ratio from 15 to 3, while the critical temperature $T_c$ decreased by only 1-2 K. These results point to an attractive way to prepare MgB$_2$ for magnet applications.

A major obstacle to applying MgB$_2$ superconductors in magnets is their rather low value of irreversibility field $H^*(T)$. Typical $H^*$ values for bulk samples, ~4 T at 20 K and ~7 T at 4.2 K, lie well below those for Nb47wt.%Ti (10.5 T at 4.2 K) and (Nb,Ta)$_3$Sn (~25 T at 4.2 K), the present materials used to make high field superconducting magnets. Irreversibility fields of ~4 T at 20 K are marginal for magnet use because high values of the critical current density $J_c$ occur only below ~ $0.6H^*(T)$. On the other hand, some thin films of MgB$_2$ have $H^*(4.2$ K$)$ values approaching 20 T in perpendicular field and 40 T in parallel field.[1] Such films have very high resistivity at $T_c$, 400 $\mu\Omega$ cm, as compared to clean-limit bulk samples, 1 $\mu\Omega$ cm or less.[2,3] This suggests that films can be driven to dirty-limit superconductivity, for which the zero temperature upper critical field $H_{c2}(0)$ is

$$\mu_0 H_{c2}(0) = 3110 \, \rho \, \gamma \, T_c \quad \text{tesla}, \quad (1)$$

and where $H^*(T)$ is approximately 85% of $H_{c2}(T)$.[14] The very high normal state resistivity just above $T_c$, $\rho(40$ K$)$ for MgB$_2$, of many films strongly enhances $H^*$ even if the Sommerfeld coefficient $\gamma$ and the critical temperature $T_c$ are reduced somewhat. However, a simple application of this formula is problematic, in that lack of texture, porosity, and second phases can all interrupt the current path. Resistivity comparisons are then uncertain by a factor of ~2, making it unclear how different fabrication procedures change the actual intragrain scattering length, which is what really determines $H_{c2}$. Thus a central question for MgB$_2$, whether significant electron scattering can be developed in bulk specimens to meet or exceed $H_{c2}$ values found in thin films, is yet quite unclear.

Previous experiments have found that irradiation,[4,5] high energy ball milling,[6] and mechanical crushing during wire drawing or tape rolling,[7,8,9] all raise $H^*$, although crystallographic texture is sometimes a complicating factor. Samples prepared from prereacted MgB$_2$ have $\rho(40$ K$)$ typically >50 $\mu\Omega$ cm, while samples made from direct reaction of Mg and B can reach values as low as 0.4-1 $\mu\Omega$ cm.[3,4] In this letter, we show that it is possible to double the irreversibility field by annealing clean, bulk MgB$_2$ in Mg vapor, while $T_c$ is only slightly reduced. $\rho(40$ K$)$ also increases from 1.0 to 18 $\mu\Omega$ cm, while the resistivity ratio decreases from ~15 to ~3. These results point to a simple means of making MgB$_2$ much more suitable for magnet applications.

A low resistivity control sample (sample A) was prepared using a similar technique[10] as earlier work,[15,11] from isotopically 99.5% enriched crystalline $^{11}$B (-325 mesh, Eagle-Picher) and pure Mg. A stoichiometric mixture was put in a Ta crucible, welded under argon, sealed in a quartz tube under vacuum, then reacted at 950 °C for 24 hours, and quenched to room temperature. X-ray and neutron diffraction detected only peaks from MgB$_2$, without any indication of Mg, MgO or MgB$_4$.

Rietveld refinement of the neutron diffraction data indicated that sample A was deficient in Mg.[12] This provoked us to make two additional samples (B and C) by annealing pieces of sample A in Mg vapor. Samples B and C were separately sealed with Mg flakes in an evacuated Nb tube, the Nb crucible being sealed in evacuated quartz. To avoid contact between liquid Mg and MgB$_2$, the Mg was isolated behind partial crimps. Heat treatment took place in a horizontal furnace at 960 °C for 10 hours. Sample B was slow cooled (10 h from 960 °C to room temperature) while C was quenched.

To define the geometry for transport and magnetization experiments, bars 5.0 × 1.5 × 0.3 mm were made by grinding parallel faces with a tripod. AC electrical resistance measurements were made at a current density of ~1 A/cm$^2$ in a 9 T Quantum Design PPMS, while an Oxford 14 T vibrating sample magnetometer (VSM) and a Quantum Design SQUID magnetometer were used for magnetization and inductive $T_c$ measurements.

Sample A exhibits superconducting and normal state properties similar to those reported for other clean limit bulk samples,[3] as indicated by the normalized resistivity $\rho(T)/\rho(300$ K$)$ in Fig. 1. Sample A has a resistivity ratio $RR = \rho(300)/\rho(40) \approx 14.7$ and $\rho(40$ K$) = 1.0$ $\mu\Omega$ cm. The superconducting transition (Fig. 1 inset) is very narrow ($\Delta T_c = 0.2$ K), with a midpoint at 39.0 K. This value is consistent with the loss of inductive shielding measured in a 5 mT field. Samples B and C exhibit a small reduction in midpoint $T_c$ to 36.9 and 37.4 K and a decrease in $RR$ to 3 as is also shown in Fig. 1. Their $\rho(40\,K)$ values are much higher, 18 $\mu\Omega$ cm (B) and 14 $\mu\Omega$ cm (C).

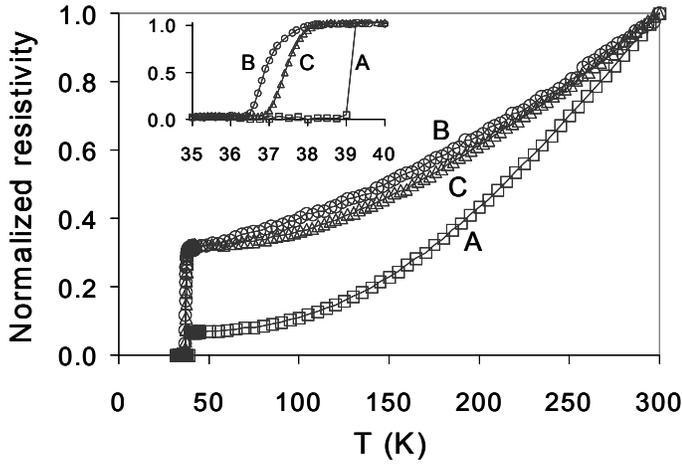

FIG. 1. Resistivity normalized to $\rho(300\,K)$ vs. temperature for sample A (squares), sample B (circles), sample C (triangles). The inset shows the transitions near $T_c$ normalized to $\rho(40\,K)$.

In-field resistive transitions are shown in Fig. 2 for all three samples. Reaction with Mg vapor makes large changes to the transitions, markedly shifting the onset of resistance to higher fields, narrowing the transition width, and significantly reducing the strong magnetoresistance observed in sample A. If we define $H^*$ by the $H(T)$ values corresponding to 10% of the normal-state resistivity, then the point at which $H^* = 9$ T occurs at 10 K for sample A, while this is increased to ~24 K for samples B and C. This implies more than a doubling of the slope of $H^*(T)$ for B and C, since the $T_c$ values of all 3 samples are comparable. We note that the magnetoresistance is small for sample B and absent in C, but otherwise the samples have similar enhancements of $H^*$.

Critical current density $J_c$ was determined from the magnetization hysteresis loops using the appropriate critical state model,[13] as shown in Fig. 3. The $J_c$ values of sample A are rather low, being below $10^5$ A/cm$^2$ in fields above 2 T at 4.2 K, which is consistent with weak grain boundary flux pinning (due to the large, ~2 $\mu$m grain size) and weak intragrain pinning when the coherence length is long. Samples B and C exhibit comparable $J_c$ values at low fields, and possess much less field dependence than for A. Since the Kramer function is not a good fit for any data set, we apply an empirical definition of the irreversibility field $H^*(T)$, namely the field at which $J_c$ falls to 10 A/cm$^2$, about a factor of 5 above the resolution limit of the VSM. This definition is not ambiguous because $J_c$ falls steeply as $H$ increases. $\mu_0 H^*$ rises from 4.2 T (sample A) to ~6 T (B and C) at 20 K and from 7.4 T (A) to 14.5 T (B) and 11.8 T (C) at 4.2 K. The 14.5 T irreversibility field of sample B is among the highest values obtained magnetically for bulk

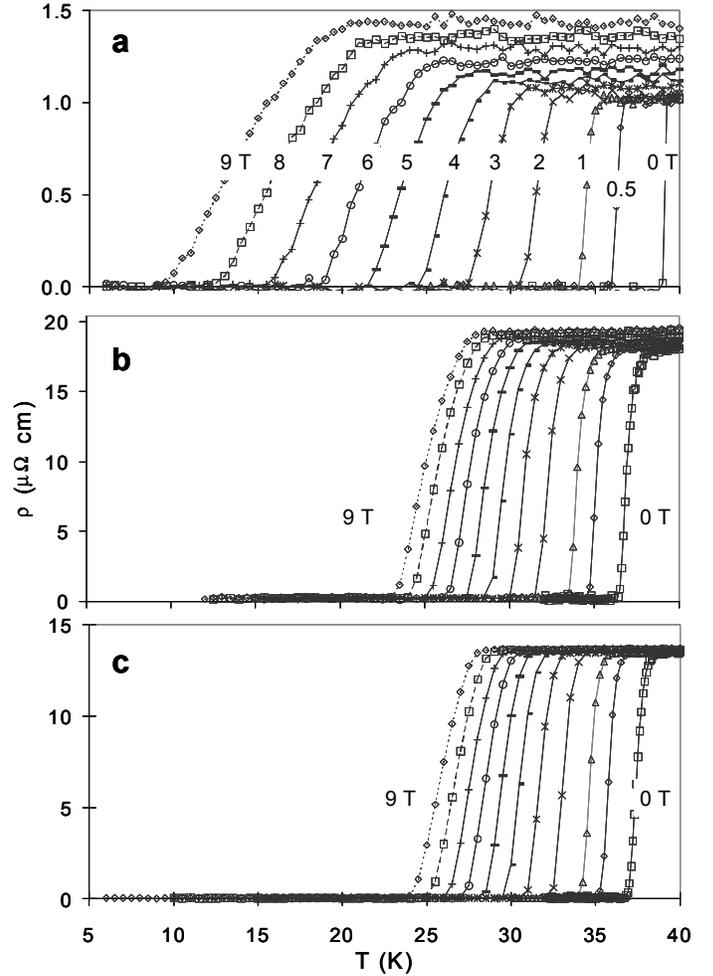

FIG. 2. Resistive transitions of sample A (a), B (b) and C (c) in applied magnetic fields up to 9 T, made at a measuring current density ~1 A/cm$^2$. The 10% of $\rho(40\,K)$ resistivity points are plotted as the irreversibility fields in Fig. 4.

MgB$_2$ at 4.2 K. Magnetization $J_c$ values of Gümbel et al.[19] at 20 K fell below 10 A/cm$^2$ at about 6.1 T, slightly higher than for our samples. To check for possible effects of enhanced $H^*$ due to texture, sample B was measured in two orthogonal orientations but no difference was seen. Fig. 4 compares $H^*(T)$ defined when the resistivity is 10% of the $\rho(40\,K)$ value. Again, $H^*(T)$ rises by a factor of ~2 for samples B and C compared to sample A. By extrapolating the plots in Fig. 4, $H^*(20\,K)$ increases from 6 T for sample A to ~13 and ~14 T for samples B and C. We also plot onset (i.e. $\rho = 0$) values for the ball-milled, nanocrystalline sample of ref. 19 and the 10% of $\rho(40\,K)$ values for film 1 of ref. 14 with $H$ perpendicular to the film plane. Taken as a whole, Fig. 4 depicts a clear trend of resistivity, critical temperature, and the slope $dH^*/dT$. As $\rho(40\,K)$ increases from 1 to 360 $\mu\Omega$ cm, $T_c$ is reduced from 39 to 31 K, while, $dH^*/dT$ increases from ~0.3 to ~0.8 T/K. We note that the $H^*(T)$ slopes for samples B and C are about the same as for the ball-milled sample, but samples B and C have superior properties because $T_c$ is above 37 K. This property combination for B and C produces the high extrapolated value for $\mu_0 H^*$ of 13-14 T at 20 K. Furthermore, the somewhat lower resistivity of samples B and C, as compared to the ball-milled and the thin film samples, indicates that further improvement in $H^*$ can still be obtained.

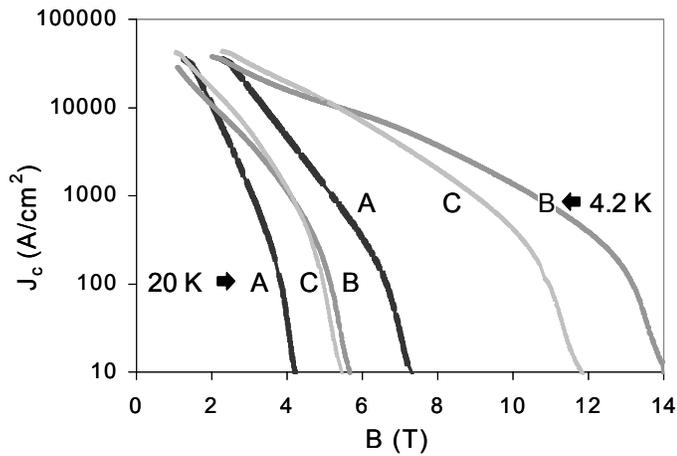

FIG. 3. Critical current density calculated from M-H loops for samples A, B, C at 4.2 K and 20 K. The magnetic irreversibility fields are determined at 10 A/cm$^2$.

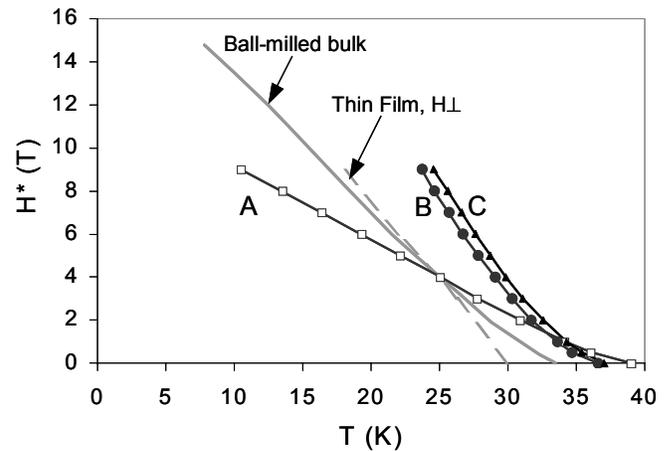

FIG. 4. Temperature dependence of the irreversibility line as determined by small-current transport measurements for MgB$_2$ samples from various sources. $H^*$ is defined when the resistivity is 10% of the value at 40 K. Data from sample A (squares), B (circles) and C (triangles), are compared with data from ref. 19 on a ball-milled nanocrystalline sample (solid line) and from ref. 14 on a c-axis textured thin film (film 1) with $H$ perpendicular to the film plane (dotted line).

An important conclusion of these analyses is that, whether the irreversibility field is defined by magnetic or by transport criteria, samples B and C show a factor of ~2 increase over $H^*(T)$ of sample A. This is significant because $H^*(T)$ values for clean limit sample A are consistent with those measured by Finnemore et al.[15] for low resistivity MgB$_2$.

Comparison of Fig. 3 and Fig. 4 underscores the difficulty of comparing irreversibility fields from different sources and measurement methods in porous bulk samples of an anisotropic superconductor. Valid comparisons are vital if superconducting property changes are to be correlated with changes in alloying via eq. (1). Neither our magnetization $H^*(T)$ data nor that of ref. 6 agree well with resistive transition transport data, the $H^*(T)$ curve determined magnetically lying a factor of ~2 below the corresponding transport curve, as seen elsewhere too[16]. By contrast, there is good agreement for textured thin films.[14] This suggests that the more stringent extrapolations of magnetization data in Fig. 3 better avoid artifacts due to percolation through an untextured, porous polycrystal at low current density.

In summary, we have shown that alloying clean-limit MgB$_2$ bulk samples in Mg vapor results in a significant increase of $H^*$ and $H_{c2}$. Both transport and magnetic measurements showed enhanced upper critical fields and irreversibility lines. More stringent magnetic measurements showed $H^* >14$ T at 4.2 K while transport measurements indicated that $H^*$ could reach 12 T at 20 K. $T_c$ is slightly decreased and the resistivity is increased from 1 to ~18 μΩ cm at 40 K for the alloyed sample. Comparison with other experiments suggested that the irreversibility field enhancement seen is still well short of the potential limit observed in thin films.

We acknowledge discussions and assistance from E. Hellstrom, B. Senkowicz, J. Fournelle, J. Jiang, A. Polyanskii, A. Squitieri, E. Bellingeri, and G. Grasso. This work was supported by the National Science Foundation through the MRSEC on Nanostructured Materials and the US Department of Energy. V.B. thanks the University of Genova for financial assistance.